\documentclass[12 pt]{article}

\usepackage[hidelinks]{hyperref}
\usepackage[utf8]{inputenc}
\usepackage[english]{babel}
\usepackage{authblk}
\usepackage{enumitem}
\usepackage{mathtools}
\usepackage{amssymb}
\usepackage{booktabs}
\usepackage{amsmath}
\usepackage{wrapfig}
\usepackage{tikz}
\usetikzlibrary{arrows,decorations.markings}
\usetikzlibrary{positioning}
\usepackage{amsthm}
\usepackage{dsfont}
\usepackage{setspace}
\usepackage[authoryear]{natbib}
\usepackage{multirow,array}

\usepackage{algpseudocode}
\usepackage[linesnumbered,boxruled,lined,noend]{algorithm2e}

\usepackage{bbm}
\usepackage{graphicx}
\usepackage{array, makecell}
\usepackage{float}

\usepackage{booktabs}

\usepackage[labelfont=sf,hypcap=false]{caption}

\usepackage{todonotes}

\usepackage{setspace}

\usepackage{arxiv}

\usepackage{xcolor}

\newcommand{\new}[1]{{\color{black}#1}}

\newcommand{\cxi}{{\mathfrak{c}}}

\newcommand{\mfp}{{\mathfrak{p}}}
\newcommand{\mfL}{{\mathfrak{L}}}
\newcommand{\mfN}{{\mathfrak{N}}}
\newcommand{\mfR}{{\mathfrak{R}}}
\newcommand{\mfT}{{\mathfrak{T}}}

\newcommand{\ba}{\mathbf{a}}

\newcommand{\bA}{\mathbf{A}}
\newcommand{\bhA}{{\mathbf{A}}}
\newcommand{\bQ}{\mathbf{Q}}
\newcommand{\bV}{\mathbf{V}}
\newcommand{\br}{\mathbf{r}}
\newcommand{\bff}{\mathbf{f}}
\newcommand{\bmu}{\boldsymbol{\mu}}

\newcommand{\bpi}{{\boldsymbol{\pi}}}
\newcommand{\bfr}{\mathbf{\mathfrak{r}}}

\newcommand{\co}{\text{CO}_2\text{e}}

\newcommand{\tT}{{t\in[0,T_L]}}

\newcommand{\R}{\mathds{R}}

\newcommand{\E}{\mathbb{E}}
\renewcommand{\P}{\mathbb{P}}

\newcommand{\mcN}{\mathcal{N}}

\newcommand{\mcL}{\mathcal{L}}
\newcommand{\F}{\mathcal{F}}

\newcommand{\Scal}{\mathcal{S}}

\newcommand{\A}{\mathcal{A}}

\newcommand{\ind}{\mathds{1}}

\newcommand{\ep}{\varepsilon}

\renewcommand{\shorttitle}{Multi-Agent RL for OC Markets}
  
\title{Multi-Agent Reinforcement Learning for Greenhouse Gas Offset Credit Markets\thanks{LW acknowledges support from the Natural Sciences and Engineering Research Council of Canada via a Canadian Graduate Scholarship. SJ acknowledges partial support from the Natural Sciences and Engineering Research Council of Canada (RGPIN-2024-04317, RGPIN-2018-05705) for this work. The authors report there are no competing interests to declare.}}

\author{Liam Welsh\\
	Department of Statistical Sciences\\
	University of Toronto\\
	\texttt{\href{mailto:liam.welsh@mail.utoronto.ca}{liam.welsh@mail.utoronto.ca}} \\
	\and
	Udit Grover \\
	Department of Engineering Science\\
	University of Toronto\\
	\texttt{\href{mailto:grover.udit@icloud.com}{grover.udit@icloud.com}} \\
	\and
	Sebastian Jaimungal \\
	Department of Statistical Sciences, University of Toronto\\ 
	Oxford-Man Institute for Quantitative Finance, University of Oxford\\
	\texttt{\href{mailto:sebastian.jaimungal@utoronto.ca}{sebastian.jaimungal@utoronto.ca}} \\
}

\begin{document}
\maketitle

\begin{abstract}

    Climate change is a major threat to the future of humanity, and its impacts are being intensified by excess man-made greenhouse gas emissions. One method governments can employ to control these emissions is to provide firms with emission limits and penalize any excess emissions above the limit. Excess emissions may also be offset by firms who choose to invest in carbon reducing and capturing projects. These projects generate offset credits which can be submitted to a regulating agency to offset a firm's excess emissions, or they can be traded with other firms. In this work, we characterize the finite-agent Nash equilibrium for offset credit markets. As computing Nash equilibria is an NP-hard problem, we utilize the modern reinforcement learning technique Nash-DQN to efficiently estimate the market's Nash equilibria. We demonstrate not only the validity of employing reinforcement learning methods applied to climate themed financial markets, but also the significant financial savings emitting firms may achieve when abiding by the Nash equilibria through numerical experiments.

\end{abstract}


\newpage

\section{Introduction}\label{sec:intro}

Excess emissions are one major cause of man-made climate change. These emissions consequently are causing more severe weather events, such as forest fires, hurricanes, and floods, and displacing millions of people each year. In an attempt to limit the impacts, more governments worldwide (both at the federal level and regional level) are pricing carbon in an attempt to incentivize emissions reductions and transition to greener technologies. There are over sixty carbon pricing systems worldwide that attempt to control carbon emissions, such as carbon taxes and emission markets. Despite this, only 4\% of these systems are valuing carbon emissions at a rate that would meet the $2^\circ$ C warming limit set by the Paris Agreement~\citep{santikarn2021state}. This is because there is no unified global price of carbon~\citep{globalPriceIMF}. In an attempt to minimize this disparity, many international treaties and agreements have been made \new{to} limit the impacts of climate change, including the Montreal Protocol (1987), Kyoto Protocol (1997), the Paris Agreement (2015), and the Glasgow Climate Pact (2021). These agreements all aim to limit excess \new{greenhouse gasses (GHGs)}, provide financial incentives to implement climate friendly legislation for developing nations, and increase the development of green and renewable technologies. To better track emissions within the financial sector,~\cite{kenyon2022carbon, kenyon2023carbon} introduces the \textit{carbon equivalence principle} (CEP), which argues that the addition of a term sheet structure for financial products of carbon emissions that are a result of the asset. CEP structured products, they argue, allow for increased carbon visibility and creation of net-zero portfolios, and is found to drive further sustainability efforts.

One method meant to financially incentivize firms to limit emissions in Canada is the implementation of a new GHG \new{offset credit (OC)} market~\citep{canadaFederalCarbon, canadaGHGOC}. This market is structured similar to both renewable energy certificate (REC) and carbon cap-$\&$-trade (C$\&$T) markets, of which there is extensive literature. Seminal work by~\cite{seifert2008dynamic} characterize firm behaviour in a C$\&$T as the solution to an optimal control problem in a single period setting, which is then extended to a multi-period model in~\cite{hitzemann2018equilibrium}.~\cite{carmona2009optimal} investigates optimal behaviour in carbon markets while simultaneously shedding light on some of the issues within these markets, and in ~\cite{carmona2022mean} the authors analyses the influence that carbon taxes have on energy producers in the mean-field limit.~\cite{aid2023regulationcarbon} solve the Stackelberg game between the regulator, who wishes to reduce overall carbon emissions, and emitting firms, wherein the regulator can dynamically assign emission allowances. In REC markets, regulators provide certificates to firms who produce electricity through renewable means. These markets have also been well studied, including by~\cite{amundsen2006price} who describes price volatility models for REC markets.~\cite{shrivatsSREC} solves for a firm's optimal controls in a solar REC market, which they then extend to a mean-field setting in~\cite{Shrivats2022_mean} with endogenized REC pricing, and finally to a principal agent mean-field game in~\cite{firoozi2021principal}. The influence that emissions have on investors and companies has also been well studied.~\cite{cartellier2023greenwashing} thoroughly and mathematically demonstrates why corporations employ greenwashing techniques, and the influence that environmentally conscience investors have on a firm's decision to employ such techniques. Greenwashing is a deceptive practice in which corporations exaggerate their green practices and environmental investments. Greenwashing practices are now under significant scrutiny due to international and societal pressures~\citep{nyt_greenwash}.~\cite{aldy2024GHGOCs} provides a comprehensive overview of OCs through a policy and regulation perspective, investigating the role OCs have as a compliance device, a voluntary credit for unregulated firms, and an incentive for climate conscious behaviour.

The majority of this prior research utilizes classical techniques, and there is little research on employing modern statistical machine learning, such as reinforcement learning (RL), to model these complex markets. This is burgeoning area due to both the advancements in learning theory techniques and the increasing severity of the impact that man-made climate change is having on our planet.~\cite{ahmadi2024greenhouse} use a deep RL technique to solve for the optimal trade-off between the mean-variance and the GHG footprint trade-off of a GHG emission adverse portfolio.~\cite{Tankov2024CoporateGHG} proposes machine learning methodologies to accurately estimate the direct and indirect emissions of firms using public data, allowing quantitative backed scrutiny to be applied to major polluters and firms that under-report the extent of their environmental impact.~\cite{noh2025geophysical} employ deep RL techniques for geological carbon sequestration to determine the optimal carbon injection rate for storage and to control and minimise geological noise (e.g.~seismic activity) while injecting. From the extant literature, the contemporaneous work of~\cite{groundwater} shares the most similarities to our paper and the prior work by~\cite{welshGHGOC}. There, the authors develop a finite-agent ground water market model in which agents optimize their water production with respect to their water rights, consumptions, and trading, and can realize further rewards by producing goods (e.g.~crops) using the water. The authors fully characterize and study the resulting Nash equilibrium in the single-period setting. However, due to the computational intractability, they propose using machine learning approaches to characterize the multi-period Nash equilibrium as a future direction.

In this paper, we present a framework for estimating the finite-player Nash equilibrium in GHG OC markets using a deep $Q$-learning technique called Nash-DQN~\citep{casgrain2022deep} and study the resulting equilibrium. The approach we use is computationally efficient, and through experiments and simulations we demonstrate the viability of the methodology and the financial importance for firms to actively participate in these markets. The remainder of the paper is organized as follows. In Section~\ref{sec:background}, we provide background on the Canadian GHG market structure that forms the basis of our market model.
and present the Nash-DQN methodology in Section~\ref{sec:back_nash_dqn}. The mathematical GHG OC market model and resulting RL implementation is provided in Section~\ref{sec:model} and Section~\ref{sec:algo}, respectively. Experimental results are presented in Section~\ref{sec:results}, wherein we demonstrate the versatility of the methodology. Closing thoughts and potential directions for future research are provided in Section~\ref{sec:conc}.

\section{Background}\label{sec:background}

\subsection{Market Structure}\label{sec:back_market}

Canadian emissions markets exist both provincially and federally. Provincial governments are allowed to regulate their own GHG emissions framework and market under the assumption that it meets the minimum federal pricing standards and GHG reduction targets~\citep{sadikman2022evolution}. In 2022, the Canadian federal government announced a new GHG OC system at the federal level with the goal of expanding project opportunities and streamlining the process~\citep{canadaGHGOC}. In this updated framework, there are  compliance markets and voluntary markets that firms may participate in. Firms may be regulated or unregulated, where a regulated firm is given an emissions limit over some period. Regulated firms participate in the compliance market, and if they exceed their limit are penalized proportionally to their excess emissions. To reduce this penalty, firms may submit OCs to the regulating body which they have either generated through project investment or have purchased from some other entity that generated OCs. By submitting OCs to the regulating body, firms reduce their penalty. Any unused OCs may be carried over into future years, up to an OC's expiry date. OCs generated in the voluntary market may not be used for compliance purposes, due to a less rigid regulatory procedures in obtaining generated OCs. Unregulated firms that choose to participate in the voluntary market (and are not compelled to generate OCs to offset excess emissions nor do they face any potential regulatory penalty) do so to amplify their environmental and social goals, or they may participate due to pressure from environmentally conscious investors.~\cite{tankov2023impactinvesting} demonstrates the impact that environmentally driven investors can have in curbing a firm's emissions, and further propose raising the cost of capital for more polluting firms to encourage emission reductions.

An OC is a certified financial derivative that is issued by some regulating body. Each OC represents one metric tonne (Mt) of carbon dioxide equivalent, denoted as $\co$. A $\co$ is a unit to compare different GHGs based on their warming potential by converting them to the equivalent warming potential of  CO$_2$. For example, methane has 25 times the warming potential of CO$_2$, thus one tonne of methane is equivalent to 25 tonnes of CO$_2$. Firms can acquire OCs by investing and implementing GHG reducing or removal projects, and having these projects verified by the regulating body. The amount of OCs that are produced is dependent on the project and each project has an associated cost to implement. Example projects include landfill reclamation, wetland restoration, and updating production processes. Firms must demonstrate that the proposed projects and emissions captured/reduced are outside of the scope of their usual business practices. If an OC project was found to be submitted to the regulator in bad faith, then the regulating body can invalidate any OCs that were distributed for the project and will (typically) require that these OCs be replaced through some other project investment. Hence, an OC represents a verifiable receipt that one Mt of $\co$ was captured from or prevented from entering the environment. OC generating projects in the Canadian system are verified by an independent accredited third party.

In 2022, the Canadian government set the excess GHG penalty to be \$50 per Mt$\co$, and this is scheduled to increase each year to reach \$170 in 2030, hence, providing financially incentives for firms to either reduce emissions or invest in OC generating projects. The updated federal framework consists of three key components: an OC generation regulation framework; new development protocols to quantify GHG reductions across different sectors; and an OC project registration and tracking system. Firms that participate may be regulated (i.e.~have an emissions limit) or unregulated, however regulated firms are required to participate. Unregulated firms may participate in this market, or may participate in the voluntary market, however OCs that are generated through the voluntary market may not be sold to regulated firms and used for compliance purposes.

The new Canadian federal market began accepting projects for registration in January of 2024, and as of September 2025 there are 32 registered projects and a total of 5,000 OCs issued~\citep{fedTracking}. Provincial GHG OC markets have long been established prior to this, including in British Columbia (BC) and Qu\'{e}bec. In the BC market, 9.3 million Mt of $\co$ were offset from the public sector between 2010 and 2023~\citep{offsetBC}, while in Qu\'{e}bec's 1.8 million OCs were issued between January 2014 and July 2024~\citep{offsetQUE}. Further, the Qu\'{e}bec market structure only allows 8\% of a firms emissions to be offset via OCs, hence incentivising general emission reductions~\citep{offsetQUE}. Provinces who do not have GHG markets or have markets that do not meet federal GHG pricing standards fall under the federal market.

Prior work by~\cite{welshGHGOC} characterizes the optimal behaviour in a single-player model and the mixed-strategy Nash equilibrium in a two-player game by employing classical stochastic control and game theoretic techniques coupled with finite-difference approximations. In this work, we estimate the Nash equilibrium for a finite number of regulated firms participating in the compliance market using deep learning techniques. We assume that these regulated firms all have the ability to invest in and generate their own OCs, or they can trade OCs.

\subsection{Reinforcement Learning and Nash-DQN}\label{sec:back_rl}

Obtaining finite-player Nash equilibria is, in general, an NP-complete problem~\citep{NashNP} --- i.e., it cannot be solved efficiently unless $P=NP$. Our approach is therefore to approximate such equilibria using RL techniques. RL (see~\cite{sutton2018reinforcement})  combines control theory with machine learning in order to approximate and learn optimal strategies for various problems. In such problems, an agent interacts with an environment --- without explicitly knowing its evolution, i.e., using the empirical measure --- in an attempt to learn the optimal policy by observing instances of states, actions, and rewards. In short, an agent observes an environmental state, chooses to take some action, and then receives a reward depending on the state, action, and (potentially) the next environmental state. 

Throughout this work, we work in a discrete time setting with $K+1$ equally spaced time points such that $t_0 = 0$, $t_{k+1} = t_{k} + \Delta t$, and $t_{K} = T$ and we further denote $\mfN = \{1,\dots,N\}$ to represent the set of indices of all agents. In multi-agent RL settings, each agent tries to learn their (e.g.~Nash) optimal policy through exploration with the shared environment via a Markov decision process (MDP) and react appropriately to the actions of other agents. The MDP is represented by the tuple $(\Scal,\,\A,\,\bfr,\,\P)$ where $\theta\in\Scal$ is an element in the state space $\Scal$, $\ba = \left\{a_i \right\}_{i\in\mfN} $ is the collection of agent actions where $a_i\in\A$ is a permissible action in the action space $\A$, $\br_{t_k}=\bfr(\theta_{t_k}, \ba_{t_k}, \theta_{t_{k+1}})$ is a reward function mapping to $\R^N$ where $\bfr = \left\{\bfr_i \right\}_{i\in\mfN}$ is the collection of agent reward functions and $\br_{t_k} = \left\{r_{{t_k},i} \right\}_{i\in\mfN}$ is the collection of evaluated agent reward functions, and $\P$ represents the probability measure that induces the transition probabilities $\P(\theta_{t_{k+1}} = \theta'\,|\,\theta_{t_k} =\theta,\ba_{t_k} = \ba)$ --- this measure is unknown to all agents. We denote $p(\cdot|\theta,\ba) = \P(\theta_{t_{k+1}} = \cdot\,|\,\theta_{t_k} =\theta,\ba_{t_k} = \ba)$ to be the distribution of the future state given the current state $\theta$ and agent actions $\ba$. Here, we assume that the action space $\A$ is shared by all the agents, but the techniques we present hold if action spaces are agent specific as well. At state $\theta_t\in\Scal$, each agent-$i$ takes an action $a_{{t_k},i}\in\A$, moves forward to state $\theta_{t_{k+1}}\in\Scal$, and receives a reward  $r_{{t_k},i}$ (see Figure~\ref{fig:RL_diag} for a visual representation). Throughout this work, bold variables are used to represents the collection of all agents. 

\tikzstyle{theta}=[shape=circle,draw=blue,fill=blue!10]
\tikzstyle{action}=[shape=circle,draw=red,fill=red!10]
\tikzstyle{reward}=[shape=circle,draw=green,fill=green!10]
\begin{figure}[h]
    \centering
    \begin{tikzpicture}[scale=0.9,every node/.style={transform shape},minimum width=1.0cm]

    \node[] (0) at (-4,0) {};
    
    \node[theta] (theta0) at (-2,0) {$\theta_{t_k}$};

    \draw [->,dashed] (0) to (theta0);

    \node[action] (a0) at (0,2) {$\ba_{t_k}$};
    \node[reward] (r0) at (2,4) {$\br_{t_k}$};

    \draw [->] (theta0) to (a0);
    \draw [->] (a0) to (r0);
    \draw [->] (theta0) to  [out=67.5,in=160] (r0);

    \node[theta] (theta1) at (2,0) {$\theta_{t_{k+1}}$};

    \draw [->] (theta0) to  (theta1);
    \draw [->] (theta1) to  (r0);

    \node[action] (a1) at (4,2) {$\ba_{t_{k+1}}$};
    \node[reward] (r1) at (6,4) {$\br_{t_{k+1}}$};
    
    \draw [->] (theta1) to (a1);
    \draw [->] (a1) to (r1);
    \draw [->] (theta1) to  [out=67.5,in=170] (r1);

    \node[theta] (theta2) at (6,0) {$\theta_{t_{k+2}}$};

    \draw [->] (theta2) to (r1);
    \draw [->] (theta1) to  (theta2);

    \node[] (1) at (8,0) {};
    \draw [dashed,->] (theta2) to (1);
    
    \end{tikzpicture}
    
    \caption{Directed MDP graphical representation.}\label{fig:RL_diag}
\end{figure}
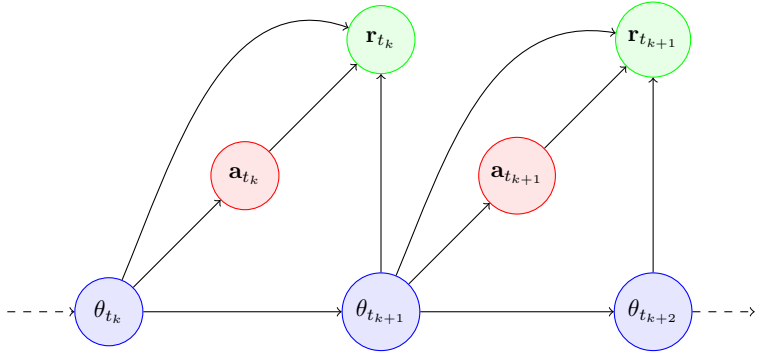

Given an MDP, agent-$i$ follows a policy $\pi_i : \Scal \rightarrow \A$ such that they select an action $a_i$ at time $t$ according to $\pi_i(\theta) = a_i$ --- we may include time in the state space to model time-dependent policies. Agent-$i$'s objective function is given by
\begin{equation}\label{eq:objective}
    R_i(\theta, \,\pi_i,\,\boldsymbol{\pi}_{-i}) = \E \left [\sum_{k=0}^{K-1} \gamma_{{t_k},i}\,{\bfr_i}(\theta_{t_k}, \ba_{t_k}, \theta_{t_{k+1}}) \right ],
\end{equation}
where $\pi_i$ is agent-$i$'s policy, $\bpi_{-i}:=\{\pi_1,\dots,\pi_{i-1},\allowbreak\pi_{i+1}\,\dots,\pi_N\}$ is the collection of other players' policies, and $\gamma_{t,i}\in(0,1]$ is a discount factor for agent-$i$,. In the sequel, we use the subscript $-i$ to denote the set $\mfN\setminus\{i\}$. An agent's $Q$-value is defined as the total (discounted) reward of an agent taking the specific action $a$ in the state $\theta$ and following the optimal policy for the remainder of the game. $Q$-learning (see~\cite{watkins1989learning,watkins1992q}) is an extremely versatile off-policy learning method for an MDP in which the learned values converge to the $Q$-values. This is accomplished with the \textit{Bellman equation}. The Bellman equation equates the value of the current state to the sum of the immediate reward and the (discounted) future value, and may be expressed in an operator form. Due to the contraction mapping of the Bellman operator, convergence to the optimal action-state value function is guaranteed in $Q$-learning under mild assumptions~\citep{watkins1992q}. In a multi-agent setting, each agent's $Q$-value is (typically) dependent on all agents' policies. For the remainder of this work, we discuss and assume a multi-agent RL setting.

A Nash-equlibria is characterized by the set of policies $\bpi^*(\theta) = \left\{\pi^*_i(\theta)\right\}_{i\in\mfN}$, such that
\begin{equation}\label{eq:nash_eq} 
    R_i(\theta,\,\pi_i,\,\boldsymbol{\pi}_{-i}^*) \leq R_i(\theta,\,\pi_i^*,\,\boldsymbol{\pi}_{-i}^*),
\end{equation}
holds for all admissible policies $\pi_i:\Scal\to\A$, for all $i\in\mfN$, and for all $\theta\in\Scal$. That is, if any agent deviates from their Nash policy while the other agents follow their Nash policy, the deviating agent is unable improve their performance. We denote $\Pi$ to be the set of admissible policies. For stationary strategies, every stochastic finite-player game has at least one Nash equilibrium in stationary strategies~\citep{fink1964}. These equilbria are, however, often difficult to compute and computationally intractable. We utilise RL techniques to efficiently approximate these Nash equilibria.
\new{We define the value function $V_i$ of the agent such that $V_i(\theta) = R_i(\theta;\,\pi_i^*,\,\boldsymbol{\pi}_{-i}^*)$, where the value function represents the sum of the current reward and (discounted) expected future rewards under all agents abiding by their equilibrium policies.}

Applying the dynamic programming principal leads to the Bellman equation for the Nash equilibria
\begin{equation}\label{eq:bellman}
    R_i(\theta;\,\pi_i^*,\,\boldsymbol{\pi}_{-i}^*) = \max_{\pi_i(\theta)\in\A}\left\{\bfr_i(\theta,\,\pi_i(\theta),\,\boldsymbol{\pi}_{-i}^*) + \gamma_i\; \E_{\theta'\sim(\cdot \vert \theta,\ba)}\left[R_i(\theta';\,\pi_i^*,\,\boldsymbol{\pi}_{-i}^*)\right] \right\}.
\end{equation}
 We define agent-$i$'s Nash state-action value function, i.e.~the $Q$-function, as
\begin{align}\label{eq:q_func_i}
    Q_i(\theta;a_i,\ba_{-i}) := \bfr_i(\theta;a_i,\ba_{-i}) + \gamma_i\;\E_{\theta'\sim p(\cdot|\theta, \ba)}\big[V_i(\theta') \big].
\end{align}
The notation $\theta'\sim p(\cdot|\theta, \ba)$ indicates that the new state $\theta'$ has distribution given by $p(\cdot|\theta,\ba)$. The $Q$-function may be interpreted as the reward agent-$i$ receives by taking the fixed action $a_i$, while others take fixed action $\ba_{-i}$, when in state $\theta$, and subsequently taking optimal actions once the new state reveals itself. Collecting the $Q$-function across agents into a vector, as well as the rewards and the value function, we re-write~\eqref{eq:q_func_i} as
\begin{align}\label{eq:q_fun}
    \bQ(\theta;\ba) = \bfr(\theta;\ba) + \gamma\;\E_{\theta'\sim p(\cdot|\theta, \ba)}\left[\bV(\theta') \right],
\end{align}
where $\bQ(\theta;\ba) := \left( Q_i(\theta;a_i,\ba_{-i}) \right)_{i\in\mfN}$ and $\bV(\theta) := \left( V_i(\theta) \right)_{i\in\mfN}$ denotes the collection of all agents' value functions.
Next, we define the Nash operator $\mcN$, which acts as follows. Let $\bff:\A^N\to \R^N$, then $\bff^*:=\mcN_{a\in\A} \bff$ is given by
$\bff^* = \bff(\ba^*)$, where $\ba^*$ satisfy
\[
f_i(a_i,\ba_{-i}^*) \le f_i(a_i^*,\ba_{-i}^*)
\]
for all $i\in\mfN$, and $f_i$ denotes the $i$-th component of $\bff$ (see~\cite{casgrain2022deep} for a formal definition) --- we assume here that there is a unique such point. The Nash operator maps a vector of functions that accept a vector of actions to the vector of values at the Nash-equilibria.
Using the Nash operator, we can relate the $Q$-functions and the value functions through the Bellman equation~\eqref{eq:bellman} as
\begin{equation}\label{eq:nash_op}
    \bV(\theta) = \mathcal{N}_{\ba\in\mathcal{A}}~ \bQ(\theta;\ba) = \mathcal{N}_{\ba\in\mathcal{A}} \left\{\bfr(\theta;\ba) + \gamma_i~ \E_{\theta'\sim(\cdot \vert \theta,\ba)}\left[\bV(\theta')\right] \right\}.
\end{equation}
This equation is the \textit{Nash-Bellman} equation~\citep{casgrain2022deep}.

One method that can be used to learn $Q$-functions (and consequently value functions) is \textit{deep $Q$-learning}, wherein deep neural networks (DNNs; in the sequel, we also use NN to denote \textit{neural network}) are used to approximate $Q$-functions. A DNN is a universal function approximator whose parameters are estimated in order to minimize some loss function. In deep $Q$-learning, a DNN is used to approximate the optimal or equilibrium strategy across all states. 

\subsubsection{Nash-DQN}\label{sec:back_nash_dqn}

To estimate the finite-player Nash equilibria for a GHG OC market, we use the Nash-DQN algorithm of~\cite{casgrain2022deep}, wherein the authors develop a computationally efficient deep $Q$-learning methodology to efficiently estimate the Nash equilibria of stochastic games. The authors accomplish this by combining the \textit{iLQG} framework from~\cite{todorov2005generalized} and~\cite{gu2016continuous} with the Nash $Q$-learning algorithm (NashQ) for general sum stochastic games of~\cite{hu2003nash}.~\cite{casgrain2022deep} employ local linear-quadratic approximations and additional concavity assumptions to efficiently estimate the Nash equilibrium. We provide an overview of the Nash-DQN methodology. In our market setting, discussed in detail in Section~\ref{sec:model}, each agent has a two-dimensional action space, and hence we accordingly define vectors and matrices with this in mind.


Nash-DQN differs from other RL approaches for computing Nash equilibria as it decomposes the $Q$-function into the sum of the value function $V$ and the advantage function $\bA$ as follows
\begin{equation}\label{eq:Q_decomp}
    {\bQ}^\alpha (\theta;\ba) = {\bV}^\alpha(\theta) + \bhA^\alpha(\theta;\ba),\qquad \forall \;\theta\in\Scal \text{ and } \ba\in\A,
\end{equation}
where $\alpha$ represents the parameters of the approximations (e.g., the weights and biases of a NN). This decomposition should be viewed as the defining equation for the advantage function (see~\cite{schulman2015high} for generalized advantage function estimation). The advantage function may be seen as the (negative) improvement that the optimal action is in state $\theta$ compared with taking the specific action $\ba$ followed by optimal actions afterwards. \new{That is, the advantage function can be viewed as providing a measure of action quality, separating the value of taking a specific action in a given state from the optimal action when in that state. By decomposing the Q-function into value and advantage components, the model can explicitly distinguish between the intrinsic value of a state and the relative benefit of selecting a particular action. Omitting this decomposition, standard Q-learning learns only the combined state–action value, making it more difficult to isolate the contribution of individual actions. Work by~\cite{tang2023va} demonstrates that directly learning value and advantage functions is advantageous to traditional Q-learning. Using advantage functions in Q-learning have also been shown to improve learning stability, and can further help reduce variance in the policy gradient estimation.} Using an advantage function in learning, in addition to the value function, mimics the actor-critic approach. In Nash-DQN, each agent's advantage function is approximated as being locally linear-quadratic with respect to the agents' actions with coefficients given as outputs of DNNs --- by locally linear-quadratic, we mean that the coefficients of the quadratic form in $\ba$ vary with state $\theta$. Specifically, in our setting where  agents select actions corresponding to both a rate of trading and a probability of generating, the advantage functions, for all $i\in\mfN$, admit the quadratic-form representation
\begin{equation}\label{eq:advantage}
    \bhA_i^\alpha(\theta;\ba) = - \begin{pmatrix} a_i - \mu_i^\alpha(\theta) \\ \ba_{-i} - \bmu_{-i}^\alpha(\theta) \end{pmatrix}^\intercal \begin{pmatrix}
        P_{11,i}^\alpha(\theta)\;\;  P_{12,i}^\alpha(\theta) \\
        P_{21,i}^\alpha(\theta)\;\; P_{22,i}^\alpha(\theta)
    \end{pmatrix} \begin{pmatrix} a_i - \mu_i^\alpha(\theta) \\ \ba_{-i} - \bmu_{-i}^\alpha(\theta) \end{pmatrix} + \left( \ba_{-i} - \bmu^\alpha_{-i} \right)^\intercal \Psi_i^\alpha (\theta)
\end{equation}
where $\mu_i^\alpha(\theta):\Scal \to \R^{2}$, $\bmu_{-i}(\theta) = \left(\mu_j^\alpha(\theta) \right)_{j\in\mfN\setminus\{i\}}$, $\Psi_i^\alpha:\Scal\to\R^{d_{-i}\times 2}$, $P_{11,i}^\alpha(\theta):\Scal\to \R^{2\times2}$, $P_{12,i}^\alpha(\theta):\Scal\to \R^{2\times d}$, $P_{21,i}^\alpha(\theta):\Scal\to \R^{d\times2}$, and $P_{22,i}^\alpha(\theta):\Scal\to \R^{d\times d}$
where $d :=2\,(N-1)$ and with the constraint that  $P_{11,i}^\alpha(\theta)$ is positive-definite $\forall \, \theta\in\Scal$. Further, without loss of generality, we set $P_{12,i}^\alpha(\theta) = \left(P_{21,i}^\alpha(\theta)\right)^\intercal$ as~\eqref{eq:advantage} is dependent only on the symmetric combination of those two blocks. The block matrix 
\begin{equation}\label{eq:P_mat}
    \text{P}^\alpha_i =  \begin{pmatrix}
        P_{11,i}^\alpha(\theta)\;\;  P_{12,i}^\alpha(\theta) \\
        P_{21,i}^\alpha(\theta)\;\; P_{22,i}^\alpha(\theta)
    \end{pmatrix}
\end{equation}
can be constructed using the output of a single DNN. Given the restriction that $P_{11,i}^\alpha(\theta)$ is required to be positive-definite,~\cite{casgrain2022deep} employ a Cholesky decomposition and instead model the lower triangular matrix, which we mimic here and hence we estimate $L_i^\alpha(\theta)\in\R^{2\times2}$ and compute $P_{11,i}=L_i^\alpha(\theta)\left(L_i^\alpha(\theta)\right)^\intercal.$ 

The form of~\eqref{eq:advantage} guarantees that ${Q}_i(\theta;a_i,\ba_{-i})$ is concave in $a_i$, for all $i\in\mfN$ and  that $\mathcal{N}_{\ba\in\mathcal{A}}~ {\bQ}(\theta;\ba)$ has a unique solution. Due to $\bhA_i$ being a quadratic form in $a_i$, with the dependence on $a_i$ being concave and by the very construction of \eqref{eq:advantage}, at the Nash equilibrium the advantage function for each agent vanishes, and we obtain 
\begin{equation}\label{eq:val_nash}
    \bV^\alpha (\theta) = \mathcal{N}_{\ba\in\mathcal{A}}~ \bQ(\theta;\ba)~~~~~~\text{and}~~~~~~\bmu^\alpha (\theta) = \text{arg}\,\mathcal{N}_{\ba\in\mathcal{A}}~ \bQ(\theta;\ba).
\end{equation}
This result may be seen by seeking for the first order conditions in $a_i$, for all $i\in\mfN$. Consequently, the Nash equilibrium is characterized by $\bV^\alpha$ and agents taking actions $\bmu^\alpha.$

With the above specifications of the advantage and value functions, we use DNNs to approximate ${{V}}_i^\alpha$, $\mu_i^\alpha$, $\text{P}_i^\alpha$, and $\Psi_i^\alpha$ for each agent-$i$, rather than approximating the each agent's $Q$-function by DNNs. As per usual in deep $Q$-learning, we optimize the DNN parameters by minimizing the mean square loss that derives from the Bellman equation: 
\begin{subequations}
\begin{align}
    \mcL_Q(\alpha) 
    &= \frac{1}{M} \sum_{m=1}^M \bigg|\bigg|  \bQ^\alpha(\theta^{(m)};\ba^{(m)}) 
    - \Big( \br^{(m)} + \gamma\,\mathcal{N}_{\ba'\in\mathcal{A}}~ \bQ(\theta^{(m)'};\ba^{(m)'}) 
    \Big)\bigg|\bigg| ^2
    \label{eq:loss_Q}
    \\
    &= \frac{1}{M} \sum_{m=1}^M \bigg|\bigg|{\bV}^\alpha(\theta^{(m)}) +  \bhA^\alpha(\theta^{(m)};\ba^{(m)}) - \br^{(m)} -  \gamma\tilde\bV(\theta^{(m)'})\bigg|\bigg| ^2,
    \label{eq:loss}
\end{align}    
\end{subequations}
where $(\theta^{(m)},\ba^{(m)},\br^{(m)},\theta^{(m)'})_{m=1}^{M}$ represents a batch of state, action, reward, and new state four-tuples obtained by following strategy $\bmu^\alpha(\theta)$ --- recall that $\br^{(m)}={\bfr}(\theta^{(m)}, a^{(m)}, \theta^{(m)'})$. In practice, we randomly sample a batch of states, use the DNNs $\bmu^\alpha$ to approximate the action to take in those, generate the next state when agents take actions, and generate the rewards. We then compute the loss in \eqref{eq:loss}, take a gradient step to improve this loss, and resample a new batch of state, actions, next state, reward, and repeat. A target value network $\tilde\bV$ is used to compute the new state's value and is described in more detail in Section~\ref{sec:algo}, along with additional algorithm details.

\section{OC Market Structure and RL Formulation}\label{sec:model}

 In this section we detail the OC market model setting, and resulting game, that we work in. The model setting is a discrete-time and finite-agent extension of the model originally presented in continuous time (with one or two agents) in~\cite{welshGHGOC}. Throughout, we work on a completed and filtered probability space $(\Omega, \F, (\F_t)_\tT,\P)$, where the OC price is denoted $S =\left(S_t\right)_\tT$, and $(\F_t)_\tT$ is the natural filtration generated by $S$. We assume the market is populated with $N$-many agents, where agents are labeled with $i\in \mfN$ and that there are $L$ many compliance periods, with compliance dates (i.e~the date on which firms must submit OCs or pay penalties for the prior period) occur on the dates in the set $\{T_l:l\in\mfL\}$, where $\mfL:=\{1,\dots,L\}$. In Section~\ref{sec:back_rl} time was discretized such that time evolved as  $t\rightarrow t+1$. In the sequel, we use a time grid such that all compliance periods are discretized into equally spaced partitions and define $\mfT:=\{t_0,t_1,\dots,t_K=T_L\}$ where $\Delta t = t_{k+1} - t_k$. OCs acquired prior to compliance date $T_l$ may be used for regulatory purposes on $T_l$. On a compliance date $T_l$, the regulator imposes a cost of
\begin{equation}\label{eq:penalty}
    G_{l,i}(x) = \mfp\,\left( \mfR_{l,i} - x \right)_+,
\end{equation}
on agent-$i$, where $\mfR_{l,i}$ represents the required inventory that agent-$i$ must hold on the $l$-th compliance date  to erase their penalty and $\mfp$ represent the penalty imposed per missing OC. The value $\mfR_{l,i}$ represents agent-$i$'s excess emissions during the $l$-th compliance period. Thus, agent-$i$ is required to submit $\mfR_{l,i}$ many OCs to completely eliminate this penalty, and we assume that $\mfR_{l,i}$ is deterministic and exogenously specified.
 
OCs may either be traded or generated. For each time $t_k\in \mfT$, agents choose their trade rate $(\nu_{t_k,i})_{i\in\mfN}$ and the probability of generating an OC  $(p_{t_k,i})_{i\in\mfN}$ for the next time period. We write agent-$i$'s actions at time $t$ as $a_{t_k,i} = (\nu_{t_k,i},p_{t_k,i})$. Agent inventories are tracked through their inventory process $X_i$, and the OC spot price process $S = \left(S_{t_k}\right)_{t_k\in\mfT}$ is governed by a discretized SDE specified in \eqref{eq:oc_price} below.

 When agent--$i$ trades at rate $\nu_{t_k,i}$ they incur a cost of $\nu_{t_k,i}\,S_{t_k} + \frac{\kappa}{2}(\nu_{t_k,i})^2$, where the first term represents the direct trading costs, and the second term represents a stylized transaction cost representing market friction where $\kappa$ represents friction parameter. Further, we impose a \textit{soft} market clearing condition through by adding an additional loss to \eqref{eq:loss}. This soft market clearing condition encourages the sum of the players trade rates to be zero --- inducing a closed market.  We impose a soft clearing condition, rather than a hard one (where, e.g., one agent is singled out as the absorbing agent), to the agents trading rates are treated symmetrically. The loss that achieves this soft clearing condition is
\begin{equation}\label{eq:trade_loss}
    \mcL_\nu(\alpha) = \frac{\varphi_j}{M}\sum_{m=1}^M\left[ \sum_{n\in\mfN} \nu_{t_k,n}^{(m)} \right]^2,
\end{equation}
where the coefficient $\varphi_j$ updates with each epoch ($j$) so that ~\eqref{eq:trade_loss} and ~\eqref{eq:loss} are of the same magnitude. More specifically,
\begin{equation}\label{eq:loss_magnitude}
    \varphi_{j+1} = (1-\phi_{\mcL})\,\varphi_j + \phi_{\mcL}\,\varphi_j\,\frac{\left(\mcL_Q\right)_j}{2\,\left(\mcL_\nu\right)_j}
\end{equation}
 where $\phi_{\mcL}\in(0,1)$, and typically closer to zero, allows for a soft update to $\varphi$, and $\left(\mcL_Q\right)_j$ and $\left(\mcL_\nu\right)_j$ are the $Q$-loss and trade loss at the $j^{th}$ epoch, defined by~\eqref{eq:loss} and~\eqref{eq:trade_loss} respectively.  

Agent-$i$ generates $\xi_i$ OCs for a predetermined price $\cxi_i$ with probability  $p_{t_k,i}$. 
If the agent generates OCs at time step $t_{k+1}$, then their inventory  updates according to $X_{t_{k+1},i} = X_{t_{k},i} + \xi_i + \nu_{t_{k},i}\,\Delta t$ and they incur an additional cost of $\cxi_i$, otherwise their inventory updates according to $X_{t_{k+1},i} = X_{t_{k},i} + \nu_{t_{k},i}\,\Delta t$. 


Generation of OCs increases their supply and therefore negatively impacts price. We model this as a downward jump in OC price by the amount $\eta\,\xi_i$. The OC trading price, $S$, is a Brownian bridge that converges to the penalty price $\mfp$ at the end of each compliance period and also incorporates the generation price impact. The OC price has the evolution
 \begin{equation}\label{eq:oc_price}
     S_{t_{k+1}} = \left(S_{t_k} - \eta\sum_{n\in\mfN} \xi_n \ind_{p_{t_k,n} > U_{t_k,n}} \right) \frac{T_l - t_{k+1}}{T_l - t_k} + \mfp\, \frac{\Delta t}{T_l - t_k} + \sigma\,\sqrt{\Delta t\frac{T_l - t_{k+1}}{T_l - t_k}}\,Z_{t_k}\;, 
 \end{equation}
 where $Z_{t_k}\sim N(0,1)$ and $U_{t_k,n}\sim\mathcal U(0,1)$, for all $t_k\in\mfT$, $t_k\in[T_{l-1},T_l)$ and $l\in\mfL$. This evolution is a discretely sampled Brownian bridge between compliance dates, where the bridge is pinned to $\mfp$ at the end of each compliance period. This pinning prevents potential arbitrage opportunities so that OCs cannot \new{be traded} at a discount or premium with respect to the penalty near \new{or on} a compliance date.
 

For our GHG OC market, the environment at time $t_k\in\mfT$ consists of 
 $\theta_{t_k} = (t_k,\,S_{t_k},\,X_{t_k,1},\dots,\allowbreak X_{t_k,N}) = (t_k, S_{t_k}, \mathbf{X}_{t_k})$, which includes  time, the OC price, and all agents' inventories. Under policy $\bpi$, Agent-$i$'s action at time $t_k$ is $a_{t_k,i}^\bpi = (\nu_{t_k,i}^\bpi,\,p_{t_k,i}^\bpi),$ a trade rate and generation probability. With this and the previously defined compliance penalty function~\eqref{eq:penalty}, we can re-write their reward function as
\begin{equation}\label{eq:rl_reward}
    \bfr_i(\theta_{t_k}, \ba_{t_k}^\bpi, \theta_{t_{k+1}}) = -\ind_{{t_{k+1}} = T_{l\in\mfL}}\,G_{l,i}(X_{{t_{k+1}}, i}^\bpi) - S_{t_k}\,\nu_{t_k,i}^\bpi - \frac{\kappa}{2}\left(\nu_{t_k,i}^\bpi\right)^2 - \mathfrak{c}_i\,\ind_{p_{t_k,i}^\bpi > U_{t_k,i}},
\end{equation}
and their criterion for a generic action policy $\bpi$ as
\begin{equation}\label{eq:gen_critetion}
    J_i^\bpi(t_k,S,\mathbf{X}) = -\E_{t_k,S,\mathbf{X}}\left[\sum_{l=1}^L \ind_{t_k \leq T_l} \;G_{l,i}(X_{T_l,i}^\bpi)+ \sum_{t_u=t_k}^{T_L-1} \left\{
    \left( S_{t_u}\,\nu_{t_u,i}^\bpi + \tfrac{\kappa}{2}\left(\nu_{t_u,i}^\bpi\right)^2\right)\Delta t + \mathfrak{c}_i\,\ind_{p_{t_u,i}^\bpi > U_{t_u,i}}
    \right\}\right],
\end{equation}
where the first term in the expectation represents the sum of future compliance penalties, the second term represents the total trading costs, the third term represents the generation costs, and $\E_{t_k,S,\mathbf{X}}[\,\cdot\,] = \E[\,\cdot \,| \,t_k,S,\mathbf{X}\,]$. We then seek the Nash equilibria where agents use the criterion~\eqref{eq:gen_critetion}. The agent's criterion in~\eqref{eq:gen_critetion} is written as a reward, hence the overall negative sign. The analysis may also be carried out using a cost criterion with only minor adjustments to the methodology.

\subsection{Implementation and Algorithm}\label{sec:algo}

In the previous sections, we introduced the Nash-DQN framework, which expresses the Q-functions $\bQ$ in terms of the value functions $\bV$ and the advantage functions $\bA$, and provided the specific market model and rewards that agents receive. In this section, we provide the algorithm for approximating these functions using DNNs and how we optimize to obtain approximate Nash equilibria. The Nash-DQN algorithm is provided in Algorithm~\ref{algo:nash_dqn}.
\begin{algorithm}[h]
	\caption{Nash-DQN algorithm for GHG OC market.}\label{algo:nash_dqn}
    \footnotesize
    \SetAlgoLined
	\KwIn{Market and agent parameters, epochs $E > 0$, batch size $M > 0$}
    Initialize DNNs with parameters $\alpha$\;
    \For{Epoch $e = 1, \ldots,E$}{
    Sample a batch of states $\{\theta^{(m)}\}_{m=1}^M$\;
    Obtain actions $\ba^{(m)} \leftarrow \bmu^\alpha(\theta^{(m)})$ with exploration given by~\eqref{eqn:exploration}\;
    Take actions and obtain new states $\theta^{(m)'}$ and rewards $\br(\theta^{(m)},\ba^{(m)},\theta^{(m)'})$ using~\eqref{eq:rl_reward_tele}\;
    Compute the loss~\eqref{eq:full_loss}\;
    Perform Adam update of network parameters $\alpha$\;
    Perform soft update of target network parameters $\tilde\alpha$ using\eqref{eq:net_update}\; 
    }
    \KwOut{Return $\alpha$}
\end{algorithm}

We use the notation $\alpha$ to denote parameters in the DNNs that approximate $\bV$ and $\bA$. Recall that each $A_i$ is parameterized by the networks 
$\mu_i^\alpha(\theta)$, $\Psi_i^\alpha$, and $\text{P}^\alpha_i$, for $i\in\mfN$. When multiple agents have identical compliance requirements, generation capacities, and generation costs they are grouped into classes. Such agents are equivalent and only differ in their state, thus, for agents within each such class we share all DNNs (i.e.~policy, advantage, and value networks). 

To obtain approximate Nash equilibria, we must minimize the total loss function (which is the sum of \eqref{eq:loss} and \eqref{eq:trade_loss})
{
\footnotesize
\begin{equation}\label{eq:full_loss}
    \mcL_j := \frac{1}{M} \sum_{m=1}^M 
    \left\{ \Big\|{\bV}^{\alpha_j}(\theta^{(m)}) +  \bhA^{\alpha_j}(\theta^{(m)};\ba^{(m)}) - \br^{(m)} -  \gamma\bV^{\tilde\alpha_j}(\theta^{(m)'})\Big\| ^2
    + \varphi_j\left[ \sum_{n\in\mfN} \nu_{t,n}^{(m)} \right]^2
    \right\},%
\end{equation}%
}%
where the subscript $j$ indicates the epoch number and $(\theta^{(m)},\ba^{(m)},\br^{(m)},\theta^{(m)'})_{m=1}^{M}$ represents a batch (of size $M$) of state, action, reward, and new state four-tuples obtained by following strategy $\bmu^{\alpha_j}(\theta)$ with exploration applied to the actions. Furthermore, $\tilde\alpha_j$ denotes the parameters of a target network, which are initialized so that $\tilde\alpha_0=\alpha_0$. the target network parameters are updated according to
\begin{equation}\label{eq:net_update}
    \tilde\alpha \leftarrow \phi_V\,{\alpha} + (1 - \phi_V)\,\tilde\alpha,
\end{equation}
for some fixed $\phi_V\ll1$ --- this is termed a soft-update.
Soft updates avoids potential overfitting and avoids forgetting previously advantageous behaviour. Using target networks for evaluating the one-step ahead value is well known to increase stability in learning and creates a more robust estimated model. 

The $i$-th agent's policy network $\pi_i$ encodes two actions: a trade rate and a generation probability. To ensure trade rates are bounded, we employ the  output activation function $x\mapsto\bar\nu\,\tanh(x)$, where $0<\bar{\nu}<\infty$. To ensure the output corresponding to generation probability is in the interval $(0,1)$, we employ a sigmoid output activation function, i.e., $x\mapsto(1+e^{-x})^{-1}$.

It is well known that exploration in RL is important for training efficiency. In our setting, we allow the agent to explore by adding Gaussian noise to the actions but then clipping to maintain their output range. More specifically, we perform the following randomization on actions
\begin{subequations}
\label{eqn:exploration}
    \begin{align}
        \nu_{t_k,i} &\leftarrow \min\Big(\bar{\nu}\,,\, \max\big(\nu_{t_k,i} + c_\nu\,\ep\,Z_\nu
        \,,\, -\bar{\nu}\big)\Big)\,,
        \\
        p_{t_k,i} &\leftarrow \min\Big(1\,,\, \max\big(p_{t_k,i} + c_p\,\ep\,Z_p\,,\, 0\big)\Big)\,,
    \end{align}    
\end{subequations}
where $Z_\nu, Z_p$ are i.i.d.~$N(0,1)$ for each agent and batch and $\ep$ decreases are training proceeds so that the agents start to exploit more and explore less.

In Algorithm~\ref{algo:nash_dqn}, environment states are randomly sampled to obtain a batch of data, including time, and only one state transition is observed per iteration. This randomization increases computational efficiency. One alternative, which is slower to converge, is to begin at $t_k = 0$ and run linearly through time. One reason why this is less efficient is that given the form of~\eqref{eq:gen_critetion}, agents are  exposed to compliance penalties only when $t^{(m)} = T_{l}-1$. For computational efficiency, we instead express the compliance date penalties using telescoping sums, thereby distributing their effect over time. To this end, we define
\begin{equation}\label{eq:pen_tele}
    g_{i}(x,x') :=  \mfp \left( \left(\mfR_{i} - x'\right)_+ - \left(\mfR_{i} - x \right)_+ \right)
\end{equation}
which denotes the partial compliance penalty between two consecutive points in time with OC inventories of $x$ and $x'$. We further simplify and assume the OC requirement is constant across all periods for any given agent --- this assumption may be relaxed with minor modifications. With~\eqref{eq:pen_tele}, we re-write the sum of all compliance penalties as
\begin{align*}
    \sum_{l\in\mfL} G_{l,i}(X_{T_l,i}) &= L\,\mfp(\mfR-X_{0,i})_+
    + \sum_{t=0}^{T_L-\Delta t} F_{t_k}\;g(X_{t_{k+1},i}, X_{t_k,i})
\end{align*}
where
\begin{equation}
F_{t_k}=
    \left\{
    \begin{array}{ll}
    L, & t_k\in[0,T_1),
    \\
    (L-1), & t_k\in[T_1,T_2),
    \\
    (L-2), & t_k\in[T_2,T_3),
    \\
    \vdots
    \\
    1, & t_k\in[T_{L-1},T_L).
    \end{array}
    \right.
\end{equation}
With the above modification, the $i$-th agent's reward can be written as
\begin{equation}\label{eq:rl_reward_tele}
    \bfr_i(\theta_{t_k}, \ba_{t_k}^\bpi, \theta_{t_{k+1}}) = -F_{t_k}\,g_{i}(X_{t_k, i},X_{t_{k+1}, i}) - S_{t_k}\,\nu_{t_k,i}^\bpi - \tfrac{\kappa}{2}\left(\nu_{t_k,i}^\bpi\right)^2 - \mathfrak{c}_i\,\ind_{p_{t_k,i}^\bpi > U_{t_k,i}}.
\end{equation}

Algorithm~\ref{algo:nash_dqn} depicts a termination scheme after a fixed number of iterations $E$, however,  other stopping criteria may be used, such as loss stability or parameter stability.

\section{Results}\label{sec:results}

In this section we provide two settings to illustrate the efficacy of our methodology. The first is a four agent setting and the second is an eight agent setting. The four agent setting represents a small market where agents share the same OC requirement but have different generation capacities, and is presented in Section~\ref{sec:four_exp}. The eight agent setting represents a large market where agents all have varying generation capacities and compliance requirements representing a more diverse market. This is presented in Section~\ref{sec:large_exp}. In both settings, market model parameters are chosen for illustrative purposes but are chosen in line with the penalty of an excess emission in the Canadian GHG OC market. Tuning market parameters to real values in this case is highly non-trivial, and would constitute a separate analysis on its own; moreover, due to the current lack of data for the Canadian GHG OC market this requires (typically) proprietary information on a firm's green investments and operations. Hence, the parameters we used are to demonstrate the methodology and yield insights regarding the market design and agent behaviour. In both settings, all agents begin at time $t=0$ with zero offset credits, we train the models using 20,000 iterations of the algorithm, and we use a batch size of 256. Once the DNNs are trained, we generate 10,000 samples paths to simulate the market and use the trained DNNs to compute agent behaviour in each sample.

\subsection{Four Agent Setting}\label{sec:four_exp}

The first setting consists of a small market with four players.  The market parameters are provided in Table~\ref{tab:four_agent_market}. The various agents' OC requirements and generation parameters are provided in Table~\ref{tab:four_agent_param}. 

\begin{table}[H]
\begin{minipage}{0.45\textwidth}
\centering
\begin{tabular}{ ccccccc } 
\toprule\toprule
  $T_l$ & Time steps & & & & & \\
 (years) &(per period) & $\kappa$ & $\eta$ & $\sigma$ & $S_0$ & $\mfp$ \\  
 \midrule
 $[1, 2]$ & 24 & 2 & 0.5 & 3 & 50  & 50 \\ 
 \bottomrule\bottomrule
\end{tabular}
\vspace{0.2cm}
\captionof{table}{Market model parameters for the four agent experiment.}
\label{tab:four_agent_market}
\end{minipage}
\hfill
\begin{minipage}{0.45\textwidth}
\centering
\begin{tabular}{ cccc } 
\toprule\toprule
 Agent & $\mfR$ & $\xi$ & $\cxi$ (\$) \\ 
 \midrule
 One & 25 & 2 & 100 \\
 Two & 25 & 1.5 & 75 \\
 Three & 25 & 1 & 50 \\
 Four & 25 & 0.5 & 25 \\
 \bottomrule\bottomrule
\end{tabular}
\vspace{0.2cm}
\captionof{table}{Agent specific parameters.}
\label{tab:four_agent_param}
\end{minipage}
\end{table}
In this setting, the marginal price of generating one OC is equal to the penalty of one excess emission for all agents, however, agents generate different amounts of OCs when they do generate, allowing us to represent large, medium, and small firms. All agents have a requirement of 25 OCs per compliance period. Agents one, two, and three are able to meet their compliance goal each period if they generate with probability one through the whole time frame (as we use $24$ time steps per compliance period), while agent four is not able to.  We use such parameters to illustrate how a small firm, represented by agent four, which does not have the capital or resources to invest in large OC generating projects, will interact in this market. If an agent does not trade or generate any OCs over both periods they incur a total penalty of $\$2,500$ (recall that all agents begin with zero inventory), and hence this is the goal that each agent must beat. Table~\ref{tab:four_agent_hyper} provides the algorithm hyperparameters that we use to train the models.
\begin{table}[H]
\centering
\begin{tabular}{ cccccccccc } 
\toprule\toprule
 Learning rate & Scheduler Step size & $\gamma$ & $\phi_V$  & $\phi_{\mcL}$ & $\varphi_0$ & Nodes & Layers \\ 
 \midrule
 0.001 & 25 & 1 & 0.05 & 0.25 & 50 & 200 & 5 \\ 
 \bottomrule\bottomrule
\end{tabular}
\vspace{0.2cm}
\caption{Nash-DQN parameters for the four agent experiment.}
\label{tab:four_agent_hyper}
\end{table}

\begin{figure}[h]
\centering
\includegraphics[height=0.25\textheight]{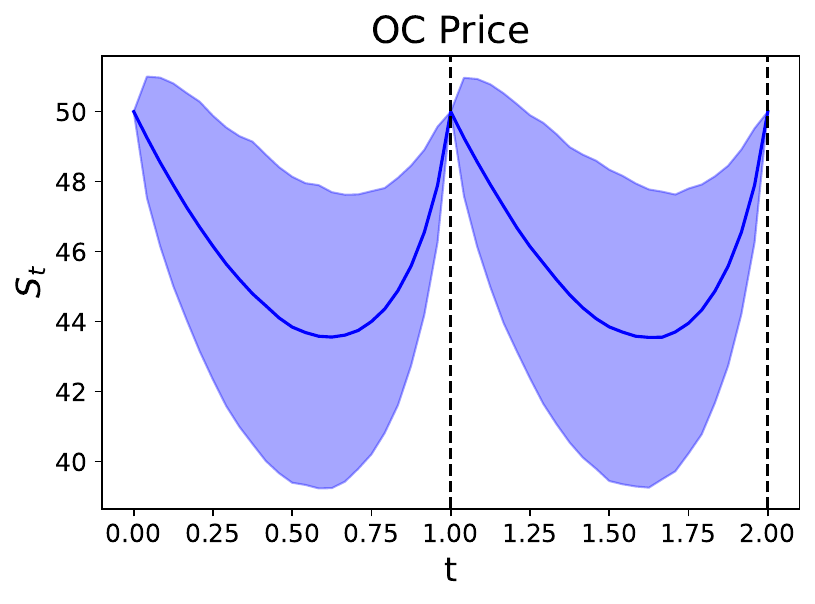}
\caption{Average OC price and 95\% quantile. Compliance dates are represented with dashed vertical lines.}
\label{fig:four_OC_price}
\end{figure}

Figure~\ref{fig:four_OC_price} shows how the OC price evolves when agents act using the trained strategies from 10,000 simulated sample paths. The shaded area is the $(5\%, 95\%)$ quantile range at each time point while the solid line is the mean price at each time point. The price evolution is pinned to the penalty value of $\$50$ at the end of each compliance period, due to the Brownian bridge dynamics, however, there is an additional downward effect due to the price impact from agents' OC generation. 

\begin{figure}[h]
\centering
\includegraphics[height=0.25\textheight]{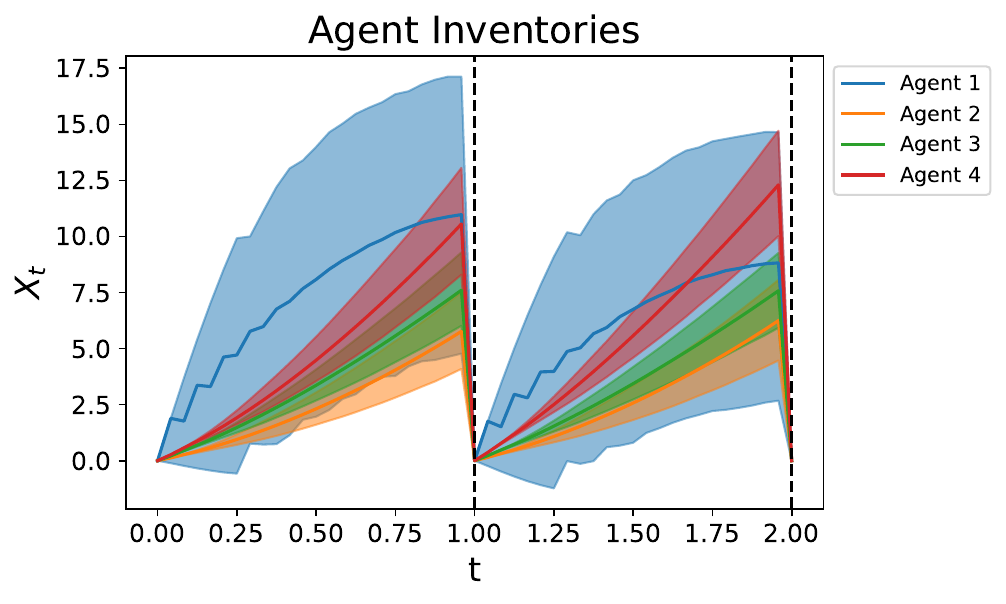}
\caption{Average agent inventory and 95\% quantile. Compliance dates are represented with dashed vertical lines.}
\label{fig:four_OC_inv}
\end{figure}

Figure~\ref{fig:four_OC_inv} shows the agents' inventories. The shaded region illustrates the quantile bands and the solid lines the average inventories. As observed, no agent fulfills the full requirement of 25 OCs in either period; consequently, all agents incur a partial penalty. \new{The sudden drop in inventory all agents experience on compliance dates is caused by agents submitting their OCs to the regulator to reduce/eliminate their compliance penalty. If an agent had an inventory exceeding their requirement at a compliance date, then they are able to carry over their excess inventory into the next period.}

\begin{figure}[h]
    \centering
    \begin{minipage}{0.475\textwidth}
        \centering
        \includegraphics[height=0.25\textheight]{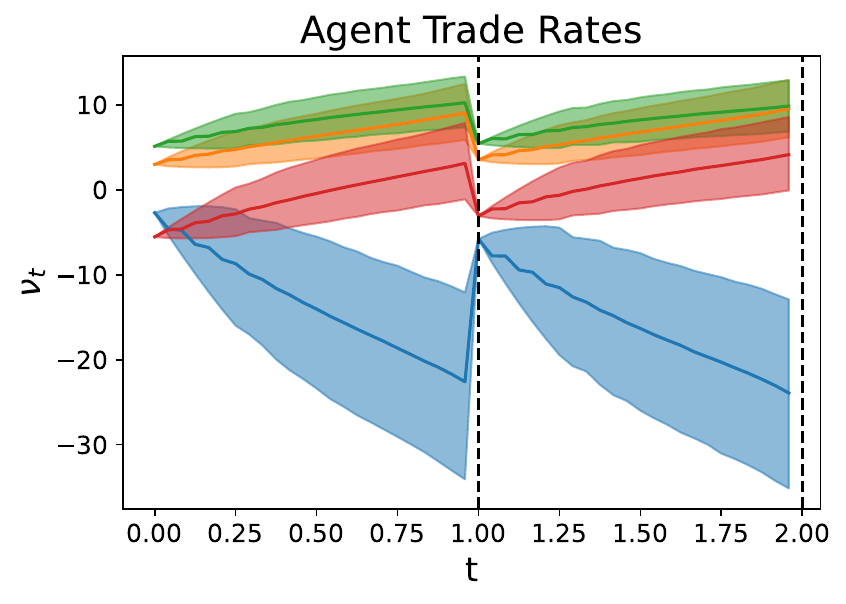} 
    \end{minipage}\vspace{0.25cm}
    \begin{minipage}{0.475\textwidth}
        \centering
        \includegraphics[height=0.25\textheight]{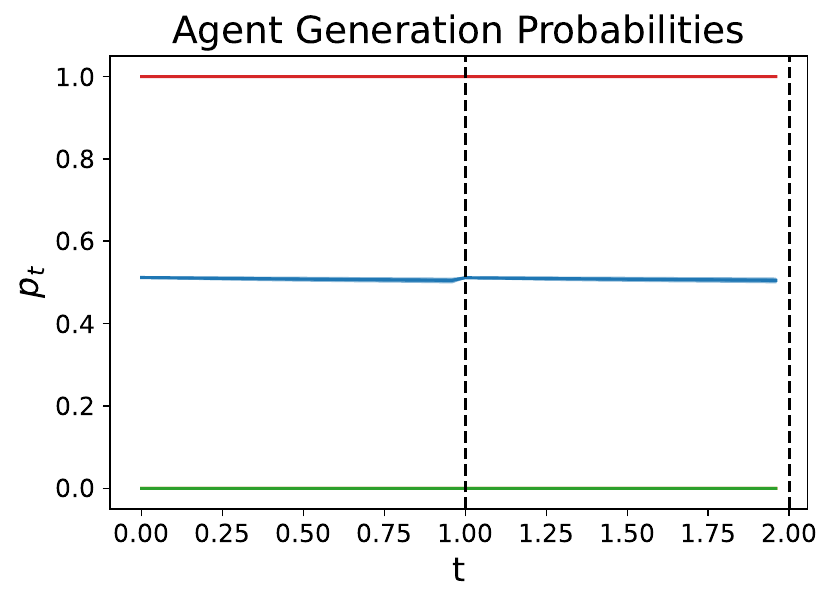} 
    \end{minipage}
    \caption{Agent trade rates and generation probabilities. Compliance dates are represented with dashed vertical lines.}
    \label{fig:four_actions}
\end{figure}

Figure~\ref{fig:four_actions} displays the agent trade rates (left panel) and generation probabilities (right panel). All agents take part in generating OCs to varying degrees, with Agents 1 (blue) and 4 (red) all generating substantially more than Agent 2 (orange) \new{and Agent 3 (green). Agent 1 generates with a probability of approximately 50\% throughout both periods, experiencing a minor increase in generation probability in the second period compared to the first period. This can be explained as the agent having more urgency to generate OCs as they enter the terminal compliance period.} Agents 2, 3, and 4 typically trade at positive rates, while Agent 1, the largest agent, takes the role of the market clearing agent, selling a partial amount of their inventory which also partially covers their cost of generating.

\begin{figure}[H]
\centering
\includegraphics[height=0.18\textheight]{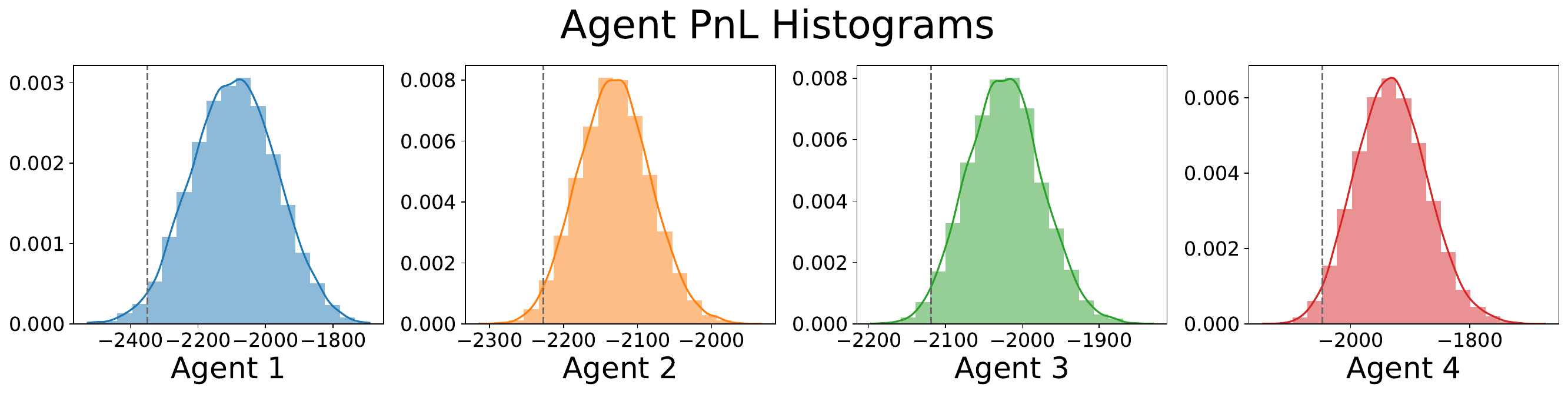}
\caption{Terminal P$\&$L histograms and kernel density estimates for each agent. Left tail expectation represented in the dashed gray line.}
\label{fig:four_PnLs}
\end{figure}

Agents' profit-$\&$-loss (P$\&$L), mean P$\&$L, left tail expectation (TE --- computed as the expected value of the lowest 5\% of terminal profits), mean sum of traded OCs, and mean sum of generated OCs are displayed in Table~\ref{tab:four_agent_PnL}. 
\begin{table}[h]
\centering
\footnotesize
\begin{tabular}{ ccccc } 
\toprule\toprule
 Agent & Mean P{$\&$}L (\$) & P$\&$L TE (\$) & Mean Traded OCs & Mean Generated OCs  \\ 
 \midrule
 One & $-2,091.73$ & $-2,349.78$ & $-28.82$ & $48.68$ \\
 Two & $-2,131.59$ & $-2,227.33$ & $12.79$ & $0.00$  \\
 Three & $-2,023.26$ & $-2,118.75$ & $16.02$ & $0.00$  \\
 Four & $-1,932.48$ & $-2,047.87$ & $0.18$ & $24.00$ \\
 \bottomrule\bottomrule
\end{tabular}
\vspace{0.2cm}
\caption{Agent mean P$\&$Ls, P$\&$L tail expectations, and mean total traded and generated OCs.}
\label{tab:four_agent_PnL}
\end{table}
Further, histograms of their P$\&$Ls are provided Figure~\ref{fig:four_PnLs}. All agents are able to surpass the benchmark of $-\$2,500$ both in their expected P$\&$L and TE, demonstrating the value of optimising and acting according to Nash equilibrium strategy. As we impose market clearing conditions with  a soft penalty term in the loss~\eqref{eq:full_loss}, the sum of the traded OC means is not identically zero, but it is small ($0.17$) relative to the amount generated ($72.68$).  A total of $72.68$ OCs were generated by all agents across both periods, hence $36.3\%$ of the $200$ total excess emissions were offset through OC generation. All agents actively participated in the market and all were found to generate OCs, conducting minor amounts of trading to offset some costs and some agents were able to take advantage of the price dip.

\subsection{Eight Player Experiment}\label{sec:large_exp}

To further demonstrate the methodology, we use a larger setting with eight agents. In this setting, some agents share generation capacities and OC requirements, hence there are agents classes. Agent classes, class population, and generation and requirement parameters are provided in Table~\ref{tab:eight_agent_param} and the market model parameters are provided in Table~\ref{tab:eight_agent_market}. As in Section~\ref{sec:four_exp}, the marginal price to generate one OC for each agent is equal to the non-compliance penalty value. Given the doubling of the number of agents, we lower the market impact of generation to  $\$0.10$ per generated OC.
\begin{table}[H]
\begin{minipage}{0.485\textwidth}
\centering
\begin{tabular}{ ccccccc } 
\toprule\toprule
$T_l$& Time steps & & & & & \\
 (years) & (per period) & $\kappa$ & $\eta$ & $\sigma$ & $S_0$ & $\mfp$ \\ 
 \midrule
 $[1, 2]$ & 24 & 5 & 0.10 & 3 & 50  & 50 \\ 
 \bottomrule\bottomrule
\end{tabular}
\vspace{0.2cm}
\captionof{table}{Market model parameters for the eight agent setting.}
\label{tab:eight_agent_market}
\label{tab:four_agent_market}
\end{minipage}
\hfill
\begin{minipage}{0.485\textwidth}
\centering
\begin{tabular}{ c c c ccc } 
\toprule\toprule
 Agent Class & Population & Label &  $\mfR$ & $\xi$ & $\cxi$ (\$) \\ 
 \midrule
 A & 2 & 1, 2 & 40 & 3 & 150 \\
 B & 1 & 3 & 30 & 2.5 & 125 \\
 C & 1 & 4 & 30 & 2 & 100 \\
 D & 2 & 5, 6 & 20 & 1.5 & 75 \\
 E & 2 & 7, 8 & 10 & 1 & 50 \\
 \bottomrule\bottomrule
\end{tabular}
\vspace{0.2cm}
\captionof{table}{Eight player setting agent parameters.}
\label{tab:eight_agent_param}
\end{minipage}
\end{table}

To ensure that agents within classes act the same when at the same point in state space, agents within a class share all DNNs. Classes are labeled from A to E, where Class A agents have the largest generation capacity and Class E agents have the smallest. The DNN hyperparameters provided in Table~\ref{tab:eight_agent_hyper}. 
\begin{table}[h]
\centering
\begin{tabular}{ cccccccccc } 
\toprule\toprule
 Learning rate & Scheduler Step size & $\gamma$ & $\phi_V$  & $\phi_{\mcL}$ & $\varphi_0$ & Nodes & Layers \\ 
 \midrule
 0.003 & 25 & 1 & 0.05 & 0.25 & 1,000 & 200 & 9 \\ 
 \bottomrule\bottomrule
\end{tabular}
\vspace{0.2cm}
\caption{Nash-DQN parameters for the eight agent setting.}
\label{tab:eight_agent_hyper}
\end{table}
Compared with the four player setting, we increase the number of layers and nodes of the DNNs so that they are better able to approximate the equilibria. As before, after training the DNNs, we generate $10,000$ sample paths to investigate how agents behave.

\begin{figure}[h]
\centering
\includegraphics[height=0.25\textheight]{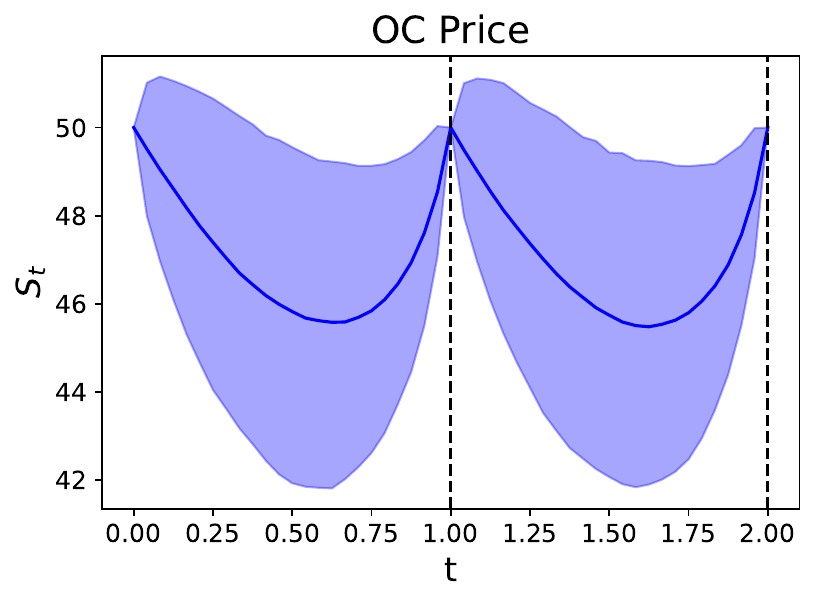}
\caption{OC price in eight player experiment.}
\label{fig:eight_OC_price}
\end{figure}

\begin{figure}[h]
\centering
\includegraphics[height=0.65\textheight]{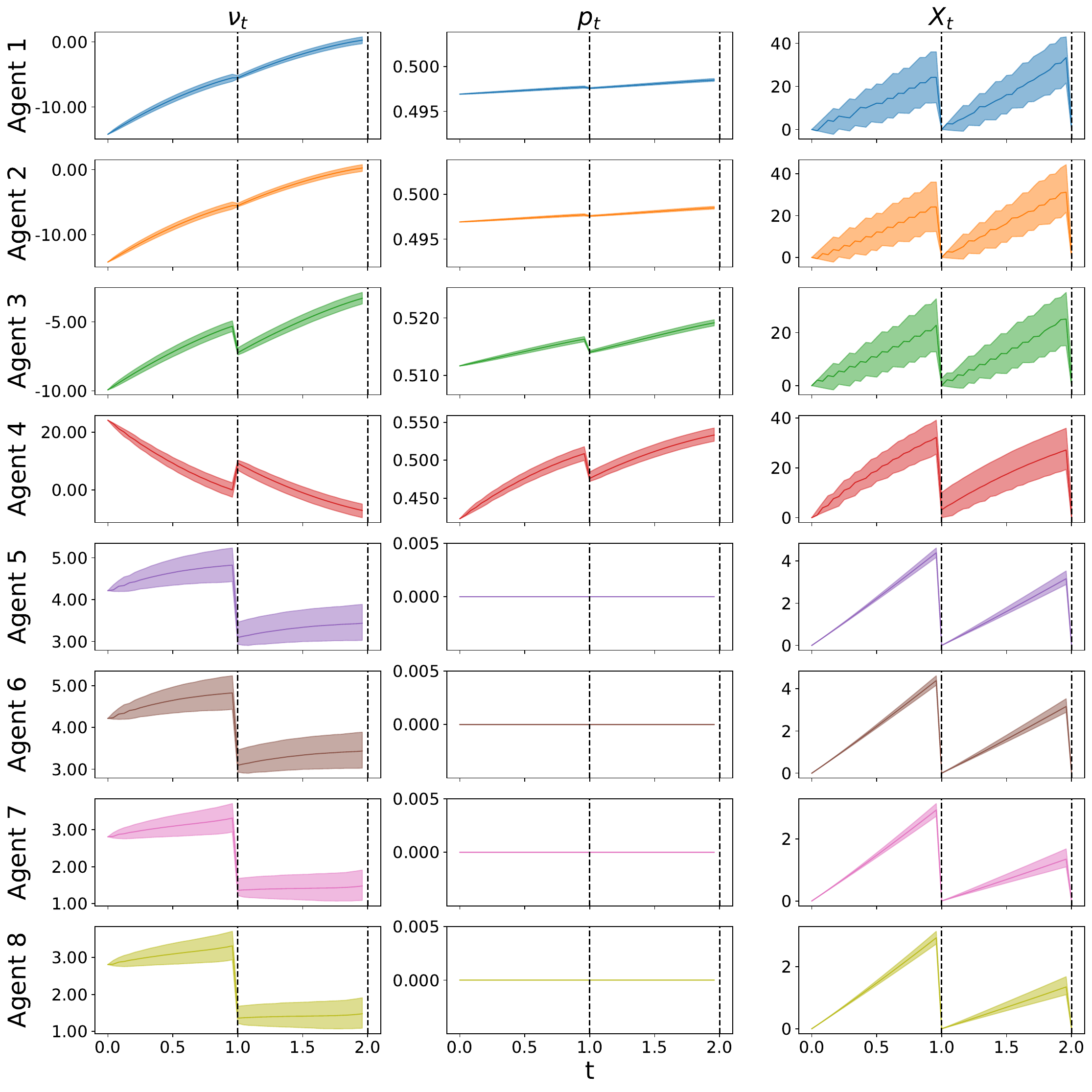}
\caption{Agents' trade right (left column), generation probability (center column), and inventory (right column).}
\label{fig:eight_indv}
\end{figure}

Figure~\ref{fig:eight_OC_price} displays the OC price quantile bands and mean path. As we saw in the four player setting, here, the effect of the pinning of OC prices at compliance dates (due to the Brownian bridge dynamics) and the impact that generation has on price are evident. Figure~\ref{fig:eight_indv} displays individual agents' trade rates, generation probabilities, and inventories. Agents within the same classes have identical results. Agents 1 through 4 have generation probabilities through both periods of approximately 50\%. These agents also have larger generation capabilities then the other agents 5 to eight, who do not generate any OCs but instead acquire their inventory through trading at positive rates.

Table~\ref{tab:eight_agent_PnL} displays the corresponding benchmarks and terminal statistics for each agent. Agent 4 is the only agent who trades both at a positive rate throughout the majority of both periods and generates OCs. All agents perform better than their benchmarks (which is the penalty assuming they do not trade or generate any OCs) and achieve non-trivial savings. Due to the soft clearing conditions, the total sum of all traded OC means is $-0.12$ and is slightly different, however, when compared with the average total generated OCs of $252.06$ this discrepancy is relatively insignificant. To offset all emissions, we require a total of $400$ OCs to be generated, hence approximately $63\%$ excess emissions are offset. This indicates a strong preference towards OC generation via project investment for the majority of agents. Further, we find that agents with the ability to generate more OCs are more inclined to generate in general, hence emphasizing the importance of being able to invest in large scale OC generating projects.
\begin{table}[h]
\centering
\footnotesize
\begin{tabular}{ ccccccc } 
\toprule\toprule
 & & & & & Mean & Mean \\
 Class & Agent & Benchmark (\$) & Mean P$\&$L (\$) & P$\&$L TE (\$) & Traded OCs & Generated OCs \\
 \midrule
 A & One & $-4,000$ & $-2,706.16$ & $-3,537.54$ & $-11.53$ & $71.61$ \\
 A & Two & $-4,000$ & $-2,709.79$ & $-3,545.89$ & $-11.53$ & $71.49$  \\
 \hline
 B & Three & $-3,000$ & $-1,910.43$ & $-2,607.97$ & $-12.47$ & $61.94$  \\
 \hline
 C & Four & $-3,000$ & $-1,694.08$ & $-2,430.13$ & $10.57$ & $47.02$ \\
 \hline
 D & Five & $-2,000$ & $-1,746.38$ & $-1,766.57$ & $7.92$ & $0.00$ \\
 D & Six & $-2,000$ & $-1,746.38$ & $-1,766.57$ & $7.92$ & $0.00$ \\
 \hline
 E & Seven & $-1,000$ & $-832.82$ & $-848.81$ & $4.50$ & $0.00$ \\
 E & Eight & $-1,000$ & $-832.82$ & $-848.81$ & $4.50$ & $0.00$ \\
 \bottomrule\bottomrule
\end{tabular}
\vspace{0.2cm}
\caption{Agent mean P$\&$Ls, P$\&$L tail expectations, and mean total traded and generated OCs two periods. Different agent classes are separated by the horizontal lines.}
\label{tab:eight_agent_PnL}
\end{table}

\section{Conclusion}\label{sec:conc}

In this work, we developed a finite-agent GHG OC model based on the Canadian market and approximated the corresponding Nash equilibrium using the deep RL methodology Nash-DQN. We applied our methodology to a four player and an eight player settings, and in doing so we  demonstrated that regulated firms who actively participate in the market (e.g.~either through trading and/or generating OCs) achieve non-trivial financial savings when abiding by the Nash equilibrium strategy. Performance is measured with respect to agents' mean PnLs and TEs against their benchmarks (i.e.~full penalty value). As more agents were added to the market model, we found that a greater percentage of OCs were generated, with respect to the total amount required to offset every agents' penalty. \new{These experiments demonstrate to regulators that the current market setting provides a fertile environment for firms to invest in OC generation, thus capturing atmospheric $\co$ and reducing overall emissions. While some firms actively generate OCs and others do not, this represents the broad spectrum of firm behaviour in the market, with some companies placing higher emphasis on environmental governance than others. Regulators may wish to create a less hospitable market environment for non-OC generating firms, and may experiment with design using the Nash-DQN framework we present.} While computing a Nash equilibrium is an NP-hard problem, we are able to efficiently approximate it in our finite-agent settings. Further computational savings are achieved by allowing agents to share DNNs when generation capabilities and OC requirements are identical. In our methodology we also include a soft market clearing condition through an additional term on the loss function, thus creating a more realistic market environment.


The methodology and analysis here can aid regulators to understand how regulated firms may behave in these markets and can lead to legislative improvements in the market \new{by providing a computational and experimental framework in which new policies and market designs can be effectively tested. The adaptability of the Nash-DQN framework will allow regulators to efficiently test various market perturbations. By analyzing firm behaviour across many experimental settings, regulators can update the market in ways to increase OC generation, OC liquidity, and/or market participation.} Given the drastically increasing and recurring consequences of man-made climate change, the importance for firms to participate in GHG OC markets and either reduce GHG emissions or implement GHG capturing projects is vital to reduce these consequences. By employing modern statistical learning techniques to solve for the optimal policies we are able to demonstrate the financial viability of active participation in GHG OC markets for firms, as well as the societal benefits these markets have in mitigating the impacts of climate change.

Both climate finance and RL (and more generally machine learning) are flourishing areas of research, hence there are many open problems that intersect the two. Within the current  framework, there remain open problems that are worthwhile investigating. First, this paper's goal was to illustrate the viability of deploying Nash-DQN in this offset credit marketing setting, and we did not calibrate our model to real data. This is because the updated Canadian market is still verifying submitted projects and no OCs have been generated, and further calibrating model parameters in a highly non-trivial problem that requires (often proprietary) knowledge of a firm's internal finances. Future work can bridge this gap by calibrating and tuning our model to real data, once it is available. Another potential avenue to take is to create a principal agent game version of our market model, such that the principal agent takes the place of a market regulator who has their own goals which the other agents (e.g.~firms) must adapt to. \cite{campbell2021deep} and~\cite{principalRL} have demonstrated this is a viable area of study, developing a scalable framework using contract theory and RL methods. Another worthwhile pursuit is to introduce stochastic OC requirements into the model. Currently, we assume the agents' requirements are deterministic and exogenous, which may not be true in the real world. Extending the methodology to include stochastic requirements would provide further complexity and realism to the model. Finally, while our model includes price impact, the OC price is given exogenously. Endogenizing the OC price within the current setting would be a very interesting avenue to explore.  Given the lack of extant literature for price endogenization in model-free RL, this has the potential to be adapted into many other types of models.


\bibliographystyle{apalike}
\bibliography{refs.bib}


\appendix

\end{document}